\documentclass{aa}  

\usepackage{graphicx}
\usepackage{subfigure}
\usepackage{txfonts}
\usepackage{verbatim}
\usepackage{float}
\usepackage{pifont}
\usepackage{tabularx,adjustbox,booktabs}
\begin{document}

\title{On the recent parametric determination of an asteroseismological model for the DBV star KIC 08626021}

\author{Francisco C. De Ger\'onimo\inst{1,2}, Tiara Battich\inst{1,2}, Marcelo M. Miller Bertolami\inst{1,2}, Leandro G. Althaus\inst{1,2} and Alejandro H. C\'orsico\inst{1,2}}
\institute{$^1$Grupo de
  Evoluci\'on Estelar y Pulsaciones. Facultad de Ciencias
  Astron\'omicas y Geof\'{\i}sicas, Universidad Nacional de La Plata,
  Paseo del Bosque s/n, (1900) La Plata,
  Argentina\\
  $^2$Instituto de Astrof\'{\i}sica La Plata, IALP (CCT La Plata),
  CONICET-UNLP\\
  \email{fdegeronimo;tbattich;mmiller;althaus;acorsico@fcaglp.unlp.edu.ar}}

\date{Received ; accepted }

\abstract{Asteroseismology of white dwarf stars is a powerful
  tool that allows to reveal the hidden chemical structure of white
  dwarfs and infer details about their present and past evolution by
  comparing the observed periods with those obtained from appropriate
  stellar models. A recent asteroseismological study has reproduced
  the period spectrum of the helium rich pulsating white dwarf KIC
  08626021 with an unprecedented precision of
  ($P_{\rm obs}-P_{\rm model})/P_{\rm model}<10^{-8}$. The chemical
  structure derived from that asteroseismological analysis is notably
  different from that expected for a white dwarf according to
  currently accepted formation channels, thus posing a challenge to
  the theory of stellar evolution.}{ We explore the relevant micro-
  and macro-physics processes acting during the formation and
  evolution of KIC 08626021 that could lead to a chemical structure
  similar to that found through asteroseismology. We quantify to which
  extent is necessary to modify the physical processes that shapes the
  chemical structure, in order to reproduce the most important
  features of the asteroseismic model.}{We model the previous
  evolution of KIC 08626021 by exploring specific changes in the
  $^{12}$C$(\alpha,\gamma)^{16}$O reaction rate, screening processes,
  microscopic diffusion, as well as convective boundary mixing during
  core-He burning.}
  {We find that, in order to reproduce the core chemical profile derived
  for KIC 0862602, the $^{12}$C+$\alpha$ nuclear reaction rate has to
  be increased by a factor of $\sim 10$ during the helium-core
  burning, and reduced by a factor of $\sim 1000$ during the following
  helium-shell burning, as compared with the standard predictions for
  this rate. In addition, the main chemical structures derived for KIC
  0862602, such as the very thin helium-pure envelope, the mass of the
  carbon-oxygen core, and the presence of a pure C buffer cannot be
  reconciled with our present knowledge of white dwarf formation. }
 { We find that within our current understanding of white dwarf
  formation and evolution, it is difficult to reproduce the most
  important asteroseismologically-derived features of the chemical
  structure of KIC 08626021.}

\keywords{stars  ---  pulsations   ---  stars:  interiors  ---  stars: 
          evolution --- stars: white dwarfs}
\authorrunning{De Gerónimo et al.}
\titlerunning{Recent asteroseismic determination of the structure of KIC 08626021}
\maketitle

\section{Introduction}


White dwarf (WD) stars constitute the most common final evolutionary 
stage of low- and intermediate-mass \citep[up to $\sim 10.6\,M_{\sun}$,][]{2015ApJ...810...34W}  stars.  In average-mass WDs, the chemical constitution of the core is mostly
a mixture of $^{12}$C and $^{16}$O, plus trace elements, of which 
$^{22}$Ne is expected to be the most abundant one. This  chemical composition is the result of the  core He-burning phase (CHeB) during progenitor evolution. At advanced stages of evolution, the WD progenitor is expected to evolve to the  thermally pulsing asymptotic giant branch (TP-AGB), where the chemical composition of the outer layers of the WD is built up \citep{2010ApJ...717..897A}. This is a critical phase that will impact the evolution and pulsational properties of the emerging WD 
\citep{2017A&A...599A..21D,2018A&A...613A..46D}.

WDs exhibit pulsational instabilities at some point in their evolution. In particular, H-deficient (He-rich) pulsating WDs (or DBVs) are found to be unstable against pulsations in the effective-temperature range
$22\,000\lesssim T_{\rm eff} \lesssim 30\,000$ K. Their
multimode photometric variations are caused by non-radial, $g$-mode
pulsations of low degree with periods between 100 and 1400 s. In the single-evolution scenario, DB WD stars are believed to be formed in the very late thermal pulse (VLTP), where the progenitor star experiences its final thermal pulse on the early cooling branch, with the result that the remaining H envelope is consumed \citep{1999A&A...349L...5H,1983ApJ...264..605I,2006A&A...449..313M}.  Alternatively, some DB WDs can be formed by mergers of two WDs, either carbon-oxygen (CO)- or helium (He)-core WDs \citep{2000MNRAS.313..671S,2002MNRAS.333..121S}. 

Details of the inner chemical structure of WDs can be inferred through
the interpretation of their pulsational spectra by means of adequate
representative models (asteroseismology). This procedure constitutes a
key technique to understand the evolution of the WD progenitors
\citep[][]{2019arXiv190700115C,2008PASP..120.1043F,2008ARA&A..46..157W,2010A&ARv..18..471A}. In
addition, asteroseismological analyses of WD stars provide strong
constraints on the stellar mass, thickness of the outer envelopes,
core-chemical composition, and stellar rotation rates
\citep[e.g.,][]{2014A&A...570A.116B,2012MNRAS.420.1462R,2012A&A...541A..42C,2011ApJ...742L..16B},
and allow to study physical processes such as crystallization
\citep{1999ApJ...526..976M,2004A&A...427..923C,2013ApJ...779...58R,2019A&A...621A.100D}.

Two main approaches have been adopted for the asteroseismology of pulsating WD stars. The first is based on static stellar structures with parameterized luminosity and chemical profiles \citep{2011ApJ...742L..16B,2014ApJ...794...39B,2019ApJ...871...13B,2014IAUS..301..285G,2016ApJS..223...10G,2017A&A...598A.109G}. The second approach is based on 
stellar evolution models computed from the zero age main sequence
(ZAMS) to the WD stage \citep[see][in the case of H-rich WD, DB, and
PG1159 stars,
respectively]{2012MNRAS.420.1462R,2013ApJ...779...58R,2006A&A...454..863C,2006A&A...458..259C,2009A&A...506..835C}. In
the first approach it is allowed for the construction of very 
  dense grid of models and the exploration of chemical structures not
necessarily expected from our current understanding of stellar
evolution. The flexibility of this method allows for extremely high
precision fits and asteroseismic models, albeit not necessarily
accurate. Parameterized chemical profiles are usually mildly inspired
by stellar evolution results.  The second approach, on the other hand,
relies on the accuracy of stellar evolution theory for a restriction
of the parameter space but is usually based on coarser grids. This
prevents high precision fits, but conversely, they are expected to be
more accurate as they are informed by a mature theory like stellar
evolution. This is particularly useful in the case of WD
asteroseismology, where the number of observed independent periods is
usually small. This asteroseismological approach is, however, affected
by current uncertainties during the progenitor evolution. These
uncertainties leave their signature on the predicted pulsation
properties and asteroseismic inferences of pulsating WDs. As recently
shown in \citet{2017A&A...599A..21D,2018A&A...613A..46D}, the impact
of these uncertainties can be quantified and bounded.

Based on the parametric approach, \citet{2018Natur.554...73G} found an
asteroseismic model with an unprecedented precision in their
pulsation-period match for the DBV star KIC 08626021,  being the
  derived stellar parameters $M_{\rm WD}=0.570\pm 0.005 M_{\odot}$,
  $T_{\rm eff}=29\, 968\pm 198$ K, log
  $g=7.92\pm 0.01 \rm \, cm\, s^{-2}$. This pulsating star, located
near the blue edge of the instability strip, has been extensively
monitored by the {\it Kepler} mission, revealing eight independent
modes with periods from 143.2 s to 376.1 s
\citep{2011ApJ...736L..39O}. The precision of the fit is of less than
$1\mu$s (i.e. a relative period difference of
$P_{\rm obs}-P_{\rm model})/P_{\rm model}<10^{-8}$), well below the
observational uncertainties of $\sim 38\mu$s.  However, this finding
has been put into question by \citet{2018ApJ...867L..30T}, who showed
that the inclusion of neutrino emission, expected in young WDs and not
considered by \citet{2018Natur.554...73G}, impacts the low order
$g$-mode frequencies up to $\sim 70\mu$Hz. Additionally, the derived
structure parameters, such as a large CO core, a high central O
abundance, a well defined C-pure mantle and a thin pure-He envelope
pose a challenge to the stellar evolution predictions. This is
particularly true for the homogeneous CO-core derived by
\citet{2018Natur.554...73G}, which is much more massive
($0.45 M_\odot$) than theoretical expectations. This disagreement
between stellar evolution theory and the asteroseismological model of
\citet{2018Natur.554...73G} is surprising in view of the previous
studies by
\citet{2010A&A...524A..63V,2010ApJ...718L..97V,2011A&A...530A...3C}
and \citet{2015MNRAS.452..123C} about the size of the He-burning
core. These asteroseismological determinations found a good agreement
between the size of the He-burning convective core
($0.22$--$\,0.28 M_\odot$), that shapes the future homogeneous CO-core
of the WD, with that coming from stellar evolution
\citep[see][]{2015MNRAS.452..123C,2015MNRAS.453.2290B}.

In this paper, we will show that the main features of the chemical
structure derived for KIC 08626021 from asteroseismology can not be
reproduced in the frame of the standard evolutionary theory.  We
assess the impact of possible uncertainties during WD and progenitor
evolution, by computing the full evolution of initial star models from
the ZAMS through the CHeB and TP-AGB phases, and finally to the WD
domain. We explore several physical processes that could lead to a
chemical structure characterized by a large CO-core with high O
abundance, a C-mantle on the top of the CO-core, and a C-rich
intershell at the bottom of the very thin He envelope, as illustrated
by the asteroseismic model for KIC 08626021.  In particular, we
explore the extra-mixing processes occurring at the border of the
convective core as well as the dependence of the
$^{12}$C$(\alpha,\gamma)^{16}$O nuclear reaction rate on the
temperature during the CHeB phase. In addition, we analyze to what
extent the evolution during the TP-AGB could affects the CHe
intershell on top of the C buffer. Finally, we assess the impact that
element diffusion should inflict on the predicted chemical profile for
KIC 08626021.

This paper is organized as follow: in Sect. \ref{numerical} we
describe the main features found in the chemical structure of a WD and
their connection with the prior evolution. In Sect. \ref{sect:results}
we present the results of our computations and finally in
Sect. \ref{sect:summary} we summary our results and conclusions.

\section{Formation of the chemical structure of a WD}  
\label{numerical}

\begin{figure*}
    \includegraphics[clip, width=0.9\linewidth]{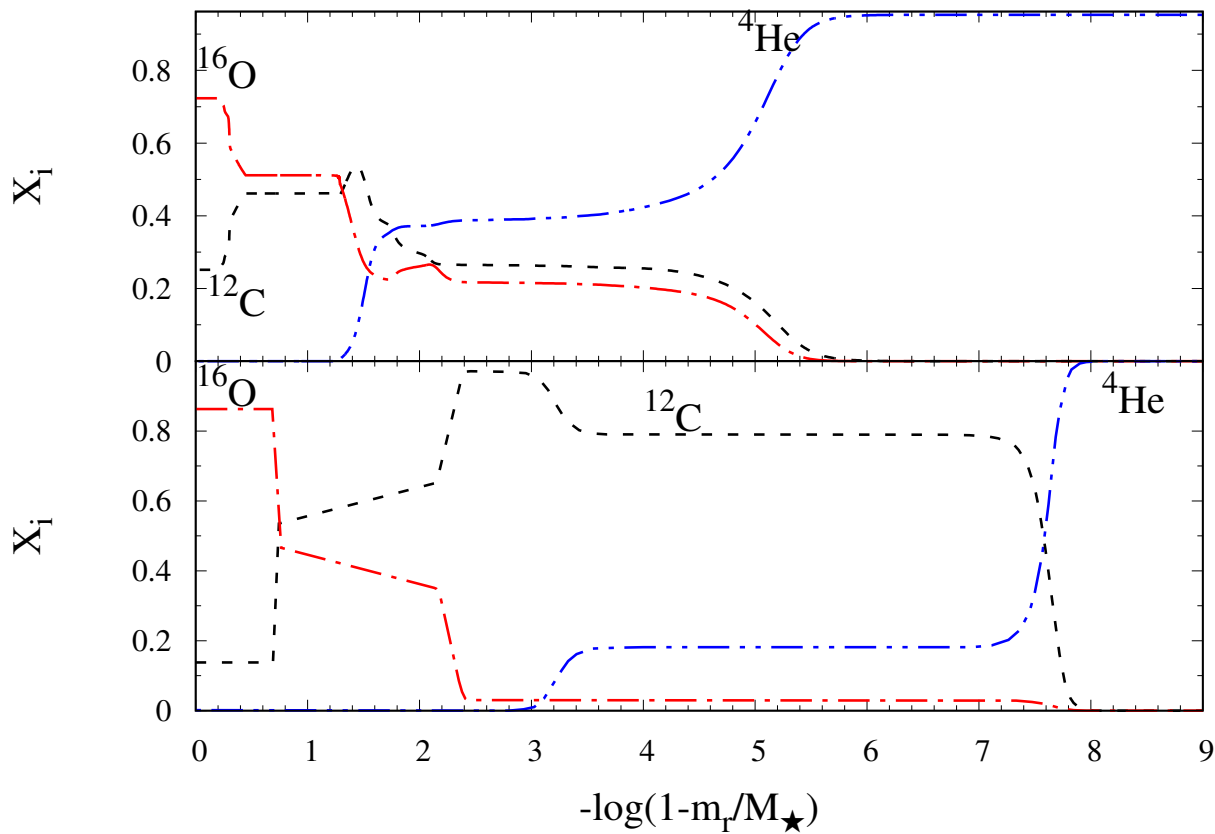}
  \caption{Upper panel: Inner distribution of O, C an He  in terms of the outer mass fraction corresponding to 
   the expectations from  a typical DBV model of mass
    $\sim$ 0.58 $M_{\odot}$ resulting from the complete progenitor evolution. Bottom panel: same as above but for 
   the asteroseismic model for the DBV KIC 08626021, \cite{2018Natur.554...73G}.} 
  \label{fig:perfil-DB-0548}
\end{figure*}

Figure \ref{fig:perfil-DB-0548} shows the typical chemical structure
of a  DBV model with similar parameters to those found by
  \citet{2018Natur.554...73G}, namely $M_{\rm WD}= 0.58M_{\odot}$,
  $T_{\rm eff}\sim 29\,000$ K and log $g=7.93$\, $\rm cm\, s^{-2}$,
derived from the full computation of the progenitor evolution (upper
panel) and the chemical-abundance profiles predicted by the
asteroseismic model for the DBV KIC 08626021 \citep[][bottom
panel]{2018Natur.554...73G}. The abundance distribution of O, C and He
from the core to the outer layers are shown in terms of the outer mass
fraction coordinate. The chemical structure of the evolutionary model
bears the clear signatures of distinct processes operative during
stellar evolution such as the CHeB, He shell burning during the AGB,
convective mixing during the TP-AGB, and element diffusion during the
WD regime.  Different regions of the WD chemical profile can be
tracked down to individual processes and, consequently, related to
specific uncertainties in stellar evolution. From center to surface,
i.e. from left to right the upper panel of
Fig. \ref{fig:perfil-DB-0548}, in brief we can identify the following:
The homogeneous central CO core [$-q \lesssim 0.3$,
$q= \log(1-m_r/M_{\star})$], which is shaped during He-core burning
and the very beginning of He-shell burning. As such, the size of the
homogeneous core and the O mass fraction are affected by uncertainties
in convective boundary mixing (CBM) and the
$\rm ^{12}C(\alpha,\gamma)^{16}O$ rate
\citep{2003ApJ...583..878S,2015MNRAS.452..123C,2015MNRAS.453.2290B,2017MNRAS.472.4900C}. Then
comes the region at $0.3 \lesssim -q\lesssim 1.5$ which is built up
during the early AGB and the TP-AGB as the He-burning shell progresses
outwards \citep{1997ApJ...486..413S, 2010ApJ...717..897A}. The details
of this region, in particular its C mass fraction and extension, are
mostly affected by CBM during the thermal pulses and at the bottom of
the convective envelope, that determine the  efficiency of third
dredge up and, in more massive stars, also the intensity of the second
dredge up. This affects the height of the C peak at $-q\sim 1.5$ which
is higher when no CBM is included. Between $1.5 \lesssim -q\lesssim 5$
come the He-C-O intershell produced during the last thermal pulse
suffered by the progenitor star. The C and O abundances in this region
are very dependent on the third dredge up history of the progenitor,
and, as such on CBM during the TP-AGB. The more efficient CBM, the
larger the final O abundance at the expense of C and He
\citep{2000A&A...360..952H, 2005ARA&A..43..435H}. The chemical
transitions at $-q\sim 1.5$ and $-q\sim 5$ are shaped by gravitational
settling, although the inner transition is far from diffusive
equilibrium when reaching the DBV instability strip
\citep{2009ApJ...704.1605A}. In addition, the total He content of the
final WD is slightly affected by details on the AGB evolution but its
order of magnitude is defined by the total mass of the final WD.

 Figure \ref{fig:perfil-DB-0548} illustrates the profound contrast between the chemical structure predicted by stellar evolutionary theory and that predicted by the asteroseismic model for the DBV KIC 08626021. In fact, both the central O abundance and, more noticeable, the extension of the CO-core are larger than 
those predicted by stellar evolution for stars with final masses $M_{\rm WD}\lesssim 0.6 M_\odot$
\citep{1997ApJ...486..413S, 2010ApJ...717..897A}.
Besides the properties of the CO core, other unconventional features are easily distinguishable in the asteroseismic model for the DBV KIC 08626021. The existence and location of the the almost pure C buffer located at $2.5\lesssim-q\lesssim 3$ is very different from that predicted by stellar evolution models. While stellar evolution models \citep[e.g.] []{2003ApJ...583..878S, 2006A&A...454..845M, 2015MNRAS.453.2290B} show a C peak formed during the late AGB evolution, its C mass fraction is always $X_C<0.8$ and it is located deeper inside the star. This last fact is connected to another unusual feature of the asteroseismic model for the DBV KIC 08626021 which is the low He content derived for that star ($M_{\rm He}=0.0001 M_{\rm WD}$), about 2 orders of magnitude lower than that predicted for WDs of average mass $\sim 0.6 M\odot$  \citep{2012MNRAS.420.1462R}. Finally, the asteroseismologically derived pure He envelope is about 3 orders of magnitude less massive than that predicted by gravitational settling at the evolutionary stage at which KIC 08626021 is found.

\section{Results }  
\label{sect:results}  

The WD evolutionary models used in this work were computed with the
{\tt LPCODE} stellar evolution code
\citep{2005A&A...435..631A,2016A&A...588A..25M}.   {\tt LPCODE}
  produces detailed WD models in a consistent way with the predictions
  of progenitor evolutionary history,  based on
  an updated physical description.  In the following we
  enumerate the most relevant physical parameters adopted in this
  work:{\it i)} Diffusive overhsooting during the evolutionary stages prior to
  the TP-AGB phase was allowed to occur following the description of
  \citet{1997A&A...324L..81H}. We adopted $f= 0.0174$ for all
  sequences, except when indicated. The occurrence of overshooting is relevant
  for the final chemical stratification of the WD
  \citep{2002ApJ...581..585P,2003ApJ...583..878S}.   {\it ii)}
  Gravitational settling and thermal and chemical diffusion were taken
  into account during the WD stage for $^1$H, $^3$He,
  $^4$He,$^{12}$C,$^{13}$C, $^{14}$N, and $^{16}$O
  \citep{2003A&A...404..593A}. {\it iii)} During the WD phase, chemical
  rehomogenization of the inner C-O profile induced by Rayleigh-Taylor
  (RT) instabilities was implemented following
  \citet{1997ApJ...486..413S}.

In the next sections we will investigate the physical processes
acting along the progenitor and WD evolution that could be responsible
of shaping the most important features of the chemical structure of
the asteroseismic model for KIC 08626021.

\subsection{Convective boundary mixing during CHeB}
\label{sect:OV}

 The treatment of CBM is one of the major uncertainties affecting
  the stellar evolutionary models and has some influence in the
  chemical profile of the WD. In particular, the incorrect application
  of the Schwarzschild criterion during the He-core burning phase can
  have a strong impact on the final chemical profile of the white
  dwarf
  \citep{2014A&A...569A..63G,2017RSOS....470192S}.
  The mass of the homogeneous central part of the CO core
  of WD models results from the interplay between convection and
  nucleosynthesis during CHeB, the ignition of the He shell at the
  very beginning of the early AGB and the late homogenization of the
  central parts driven by an inversion in the mean molecular weight of
  the stellar material \citep[see Fig. 3 of][]{1997ApJ...486..413S} .

  Interestingly, the location of the outer boundary of the convective
  core is initially governed by a self-driving mechanism
  \citep{1971Ap&SS..10..340C}. Any extension of the convective
  boundary beyond its formal value as given by the Schwarzschild
  criterion is expected to increase the C abundance of the
  neighbouring layers, thus leading to an increase in their opacity,
  and consequently $\nabla_{\rm rad}$ , and thus to a larger
  convective core. The increase of the size of the convective core
  moves the convective boundary, and CBM, even further. This process
  continues until the value of $\nabla_{\rm rad}$ equals the local
  value of the adiabatic gradient $\nabla_{\rm ad}$. Due to the
  self-driving nature of this mechanism, as soon as some mixing is
  allowed beyond the He-burning convective core, the process develops
  until it reaches its stable value. In fact,
  \citet{2007ApJ...670.1178M} showed that even atomic diffusion is
  enough to trigger this instability, eventually increasing the size
  of the convective He core. Consequently, the adoption of a bare
  Schwarzschild criterion for the determination of the convective
  borders will lead to nonphysical convective He-burning cores, where
  neutral buoyancy is not attained at both sides of the convective
  border as a consequence of the chemical discontinuity. In our case,
  this problem can either be solved by a detailed analysis of
  convective stability at both sides of the convective border
  \citep{2014A&A...569A..63G}, or by allowing for some mixing beyond
  the ill-defined convective boundary. The inclusion of even a very
  tiny CBM already allows models to grow the convective core so that
  it reaches neutral buoyancy at its outer convective boundary. Due to
  the self-driving nature of the mechanism, it is expected that the
  final size of the convective core is similar irrespective of the
  nature of the additional mixing that occurs at the convective
  boundary. A detailed account of CBM during the He-core burning stage
  of low-mass stars can be found in Section 4.2 of
  \citet{2017RSOS....470192S}.

  In addition to this self-driving mechanism, the latter He-core
  burning gives rise to the appearance of splittings in the formal
  (i.e. Schwarzschild criterion) convective core that can be modelled
  as a partially mixed region, where neutral buoyancy is attained
  \citep{1985ApJ...296..204C}. This referred to as semiconvection by
  some authors\footnote{Not to be confused with the semiconvection
    mechanism described in textbooks \citep{2013sse..book.....K} which
    is due to overstability as a consequence of non-adiabatic
    effects.}. Again, the inclusion of some minor CBM allows the
  convective zone to stay connected, and although details in the final
  chemical profiles keep a record of the exact method adopted for
  computing mixing beyond the formal convective boundary, all
  algorithms lead to similar sizes of the homogeneous central part of
  the CO core \citep{2015MNRAS.453.2290B, 2015MNRAS.452..123C}. As a
  consequence, different treatments of convective boundary mixing
  during the CHeB stage do not lead to significant discrepancies in
  the final chemical profiles of the WD, provided that some mixing is
  allowed beyond the formal Schwarzschild convective boundary.

 \begin{figure}
    \includegraphics[clip, width=1.\linewidth]{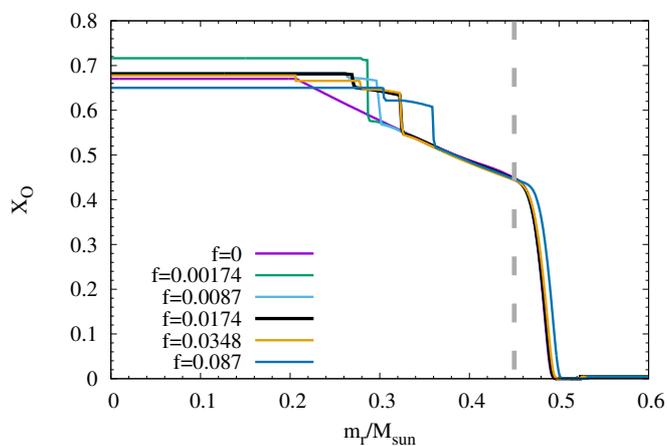}
  \caption{ Oxygen chemical profiles as a function of the mass coordinate for different assumptions of the overshooting parameter. Vertical dashed line corresponds to the extent of the homogeneous central part of the core predicted by the asteroseismic model of KIC 8626021.} 
  \label{fig:HeCB}
\end{figure}

The extent of the homogeneous central part of the core in the chemical profile derived by \citet{2018Natur.554...73G}, is about 0.45 $M_{\odot}$, much higher than the predicted by evolutionary computations, $\sim 0.32 M_{\odot}$. \citet{2018Natur.554...73G} propose that this could be due to more extra mixing during the CHeB by semiconvection or overshooting. We explore then the impact of the extension of the convective core on the final size of the homogeneous central part of the WD core. In order to do this we performed simulations starting from the same initial model ($Z= 0.01$, $M_i= 1 M_\odot$) for different values of the CBM  parameter $f$ during CHeB\footnote{The value of $f$ relates  the mixing coefficient of a layer outside the formal convective zone ($D_{\rm CBM}$) at given distance $d$ from the formal convective boundary with the mixing coefficient close to the formal convective boundary ($D_0$) via the relation $D_{\rm CBM}=D_0\times \exp{-2d/f H_P}$, where  $H_P$ is the local pressure scale height at the formal convective boundary \citep{1997A&A...324L..81H}.}. In particular, we explore values of $f= 0.00174, 0.0087, 0.0174, 0.0348, 0.087$, and $0.174$ which correspond to $1/10, 1/2, 1, 2, 5$, and $10$ times the standard value of  $f_0= 0.0174$, see \citet{2016A&A...588A..25M}. 

Fig. \ref{fig:HeCB} shows the resulting chemical profiles of our
models at the beginning of the thermally pulsing AGB phase, after the
homogenization of the central parts driven by an inversion in the mean
molecular weight of the stellar material
\citep{1997ApJ...486..413S}. As expected, as soon as some additional
mixing is allowed at the convective boundary, the size of the
homogeneous CO core is significantly enlarged. Even a very minor CBM
efficiency ($f= f_0/10$) is already enough to start the self-driving
mechanism mentioned at the beginning of this section, producing a
homogeneous CO core of $M_{\rm CO}\simeq 0.286 M_\odot$ (to be
compared with the $M_{\rm CO}\simeq 0.205 M_\odot$ resulting in the
unrealistic case in which all CBM is prevented). In comparison, further
increases in the value of $f$ by factors of 5, 10, and 20 (i. e. $f= 0.0087, 0.0174, 0.03$) lead to
relatively minor increases in the mass of the homogeneous CO core: 
$M_{\rm CO}\simeq 0.292$, 0.322, and 0.322 $M_\odot$ 
respectively.  From our previous discussion, this is an expected trend, because the main process determining the
size of the core only requires the existence of some additional
mixing, provided that it is enough to alter the layers immediately
outside the formal convective border \citep{1971Ap&SS..10..340C,
  1985ApJ...296..204C}.  Only when $f= 0.087$ is adopted, the extent of
CBM leads to a larger homogeneous core of
$0.354\,M_{\odot}$. This value is still far below the value of
$0.45\,M_{\odot}$ derived by \citet{2018Natur.554...73G} for KIC\,08626021. 
Considering that a value of $f=0.087=$ $5\times f_0$ is very high in comparison
with any calibration of the overshooting parameter, this rules out the
possibility of CBM as being the cause behind the
large CO core inferred for KIC\,08626021. Assuming a larger value of $f$, like $f=0.174$, we find  the evolution of the post-CHeB
star to be completely altered, with thermal pulses developing only
$800\,000$ yr after the end of CHeB, more than a factor 10
shorter than in a normal evolution, and thus effectively truncating
the very existence of the early AGB phase. Such a model would be
incompatible with the  existence of the early AGB phase and should
be already discarded on those grounds. And even with such
inconsistently large value of $f=0.174$, the mass of the homogeneous CO
core is reduced by the first thermal pulse to $0.386 M_\odot$ (from a
value of $0.422 M_\odot$ at the very end of the HeCB), well below the
value of $0.45\,M_{\odot}$ derived by \citet{2018Natur.554...73G}.

 The inability to produce homogeneous CO cores as large as those
reported by \citet{2018Natur.554...73G} is not a property of the
exponentially diffusive overshooting prescription adopted here but of
all studied CBM recipes. As already shown in Fig. 4 of
\citet{2003ApJ...583..878S} for standard sized WDs
($\sim 0.6M_{\odot}$) semiconvection and penetrative/mechanical
overshooting, even under extreme assumptions, lead to homogeneous CO
cores well below the value derived by \citet{2018Natur.554...73G}. A
similar result is shown in Fig. 2 of \citet{2015MNRAS.452..123C},
which in addition to penetrative overshooting and semiconvection also
explore the CO-profiles left by a moderate exponentially decaying
overshooting, and in Fig. A1 of \citet{2015MNRAS.453.2290B} which
shows the final CO-profiles under different assumptions of the
temperature gradient for the mechanical overshooting approximation
(called ``overshooting'' and ``penetrative convection'' in their work)
under the extreme assumption of a $1 H_P$ overshooting zone. In
addition to these experiments, \citet{2015MNRAS.452..123C} explored a
``maximal-overshooting'' scheme that avoids the splitting of the
He-burning core at latter stages of the CHeB phase.  This recipe leads
to slightly smaller homogeneous CO-cores than the standard exponential
and penetrative overshooting prescriptions. Finally,
\citet{2017MNRAS.472.4900C} also explored the incorporation of
Spruit’s core-growth rate
\citep{2015A&A...582L...2S}. \citet{2015A&A...582L...2S} makes
physically sounding arguments regarding the maximum rate at which a
convective He-burning core can grow in a steady regime based on the
higher buoyancy of the material ingested. This argument then sets an
upper limit to the maximum size of a He-burning core and consequently
to the size of the homogeneous CO region in the core of WDs. Fig 1 of
\citet{2017MNRAS.472.4900C} shows that Spruit’s argument also leads to
convective cores not larger than those obtained with the exponentially
decaying overshooting approximation.  All these works together show
that the outer boundary of the homogeneous CO core of a low-mass star,
like the progenitor of KIC 08626021, cannot exceed $0.35M_{\odot}$
even under the most extreme situations.

We conclude that CBM cannot make 
the homogeneous part of the core grow up to $0.45\,M_{\odot}$ without
changing drastically other parts of the stellar evolution that are
well constrained such as the existence of the early AGB phase.

\subsection{Efficiency of diffusion processes}
\label{sect:difu}
During the WD evolution several processes strongly modify the chemical structure of the progenitor star. Among them, gravitational settling is the primary shaper of WD chemical profiles, forming chemically-pure outer
layers. 
Here, we explore to what extent element diffusion processes due to gravitational settling, thermal, and chemical diffusion could be  responsible for the formation of a C-pure buffer at the top of the CO-core. We also explore for how long can a very thin He envelope survive the effects of diffusion in the absence of competing processes. Diffusion coefficients, that determine how efficient these processes are, have been calculated by various groups (e.g.  \citealt{1986ApJS...61..177P,2013PhRvL.110w5001B}). 
Differences in the diffusion coefficients are at most of one order of magnitude in the strong coupled plasma regime \citep{2013PhRvL.110w5001B,2015ApJS..220...15P}.  

\begin{figure}
  \includegraphics[width=1.\linewidth, angle=0]{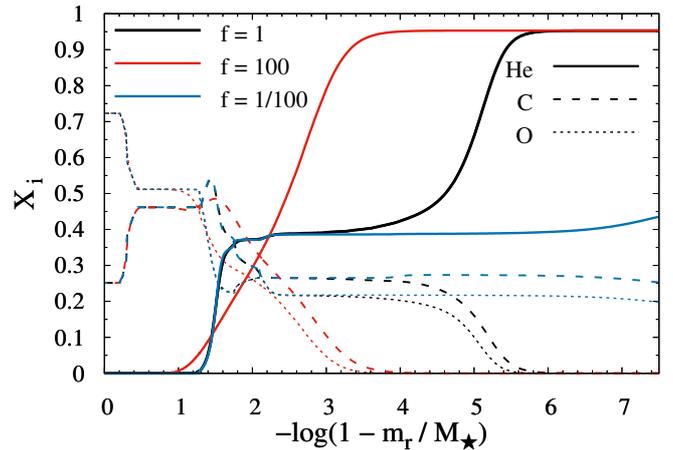}
  \caption{Chemical profiles for He, C and O of our DBV evolutionary models ($\sim$ 29000 K) in terms of the outer mass fraction, resulting from different efficiency of element
    diffusion. The values of the
    quantity $f$ indicates the multiplicative factor of the diffusion
    efficiency with respect to the standard value ($f=1$).}
  \label{fig:dif1-100}
\end{figure}
To explore the impact of time-dependent diffusion  on the chemical profile of a WD at the effective temperature and mass of KIC 8626021, we evolve a $\sim$ 0.57$M_{\odot}$ WD model from
$\sim 200000$ K to $\sim 29000$ K and modify the efficiency of the diffusion processes by a multiplicative factor $f$, $f= 0.01,\,1$ and $100$, thus widely 
covering the actual uncertainties in these processes. Fig.~\ref{fig:dif1-100} shows the chemical profiles resulting from our experiment. 
 Clearly, changing the efficiency of the diffusion processes in any reasonable amount is not expected to reproduce the main remarkable features of the asteroseimological profile of KIC 8626021. In particular, we note that the peak of C at $\log(1-m_r/M_{\star})\sim$ -1.4 does not change significantly, neither in the value of the peak nor in the position. This means that we cannot invoke diffusion as the responsible process to create an almost pure C buffer in the WD.
\begin{figure}
    \includegraphics[clip, width=1.08\linewidth]{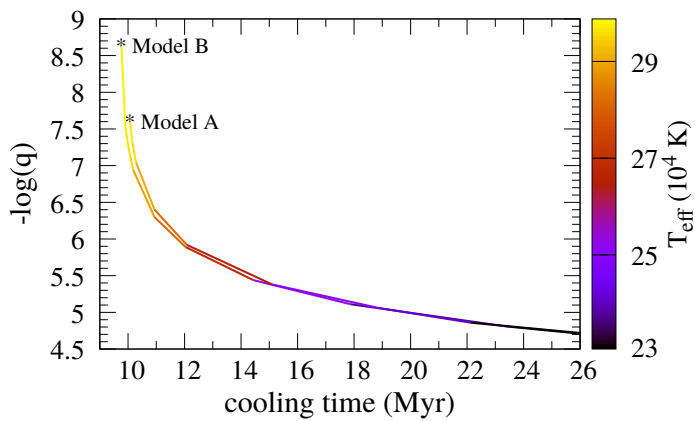}
  \caption{Time evolution of the position of the bottom of the pure-He envelope (measured in terms of the outer mass fraction $q$) from $T_{\rm eff} \sim 30000$ K, for models with initial $-\log(q)\sim 7.6$ and 8.6 (models A and B respectively). For model A (B), 0.08 (0.18) Myr is enough for diffusion processes to thicken the He envelope below $ \log(q)\sim 7.4$. } 
  \label{fig:env-evol}
\end{figure}

A difficulty also arises when trying to reproduce the thin pure He envelope derived by \citet{2018Natur.554...73G}.
Thanks to the relatively low uncertainties in the diffusion physics in
the outer regions of WDs
\citep{2013PhRvL.110w5001B,2015ApJS..220...15P} we can estimate how
long  such a thin He-pure envelope can survive.  To this end, we perform a set of numerical
experiments with {\tt LPCODE} by computing the speed of gravitational
settling at the evolutionary stage and mass of KIC 8626021. Initial
chemical profiles are  those shown in
Fig. \ref{fig:dif1-100} but with the outer He-pure envelope located at
different initial depths of $-\log(1-m_r/M_{\star})= 7.6, 8.6$ (model A and B, respectively). 
Our computations show that
in about $100000-200000$ yr the envelope becomes already thicker than the value found for KIC 8626021,
$-\log(1-m_r/M_{\star})= 7.4$ (see Fig. \ref{fig:env-evol}), falling
outside of the range of the asteroseismical solutions. These
timescales are only 1 to 2\% of the time required by standard DB WD
models to cool down to $T_{\rm eff}\sim 30000$K, which is of about 10
Myr for models in that mass range \citep{2009ApJ...704.1605A}. Hence,
if KIC 8626021 is characterized by such thin He envelope, then the WD
should have been formed by an evolutionary scenario that allowed it to
cool down to its present state about 50 to 100 times faster than
normal DB stars.

Competing processes such as strong winds or rotation could in
principle delay the action of gravitational settling. However, the
existence of strong winds in WDs is at variance with the observed
action of radiative levitation in DO stars
\citep{2018A&A...612A..62H}, which can only be effective if winds do
not prevent the action of diffusion. Also the location of the
DO-PG1159 transition \citep{2017A&A...601A...8W} can be reproduced
\citep{2000A&A...359.1042U} when WD winds decay strongly with decaying
luminosity, as expected from radiation driven wind theory
\citep[e.g. $ \dot{M} \propto L^{1.86}$ as proposed
by][]{1995A&A...299..755B}. In particular, stellar winds are expected
to stop as soon as metals sink below the photosphere and are not
available to absorb momentum from the radiation field
\citep{2000A&A...359.1042U}.  In addition, the fast drop in mass loss
with stellar luminosity proposed by \citet{1995A&A...299..755B} is
needed to provide a coherent picture of the GW Vir red edge
instability domain \citep{2012ApJ...755..128Q}. All these concerns are
reinforced by the fact that such winds would require an extreme fine
tuning of its intensity to remove almost all the initial He content
but not all.  Similarly, while rotational mixing could lead to a delay
of gravitational settling, the slow solid body rotation measured in
KIC 0826021 by \citet{2018Natur.554...73G} strongly argues against
this possibility.

\subsection{The $^{12}$C$(\alpha, \gamma)^{16}$O nuclear reaction rate and Coulomb screening}
\label{sect:co}

The chemical abundances of the CO-core, as well as of those layers
immediately above, are produced at the end of CHeB phase and the
beginning of He-shell burning.  In the previous section we show that
diffusion is unable to create the C-pure buffer, even when diffusion
coefficients beyond current uncertainties are adopted. Assuming that
diffusion is the only process able to modify the chemical structure
during WD stage, any chemical structure located so deep into the
interior of the star should be a fossil record of the previous
evolution. The O to C ratio left by He-burning is a consequence of the
competition of the 3$\alpha$ reactions that creates $^{12}$C and the
$^{12}$C$+\alpha$ reaction that destroys $^{12}$C to create
$^{16}$O. In particular, the $^{12}$C$+\alpha$ reaction is among the
most uncertain one in stellar evolution. In this section, we explore
to which extent the temperature dependence of the $^{12}$C$+\alpha$
nuclear reaction rate should be altered in order to produce the high
central O abundances together with the previously discussed C buffer.
Recently, \citet{2017A&A...599A..21D} explored the implications of the
current uncertainties in the $\rm ^{12}C+\alpha$ nuclear reaction rate
during the CHeB phase over the chemical structure and pulsation
periods of hydrogen-rich pulsating WDs. The authors found that these
uncertainties have a non-negligible impact in the chemical structure,
but as seen from their Fig. 7, it is clear that none of their models
predict the most important features of the asteroseismic model found
for KIC 08626021.

\begin{figure}
  \includegraphics[clip, width=1.0\linewidth, angle=0]{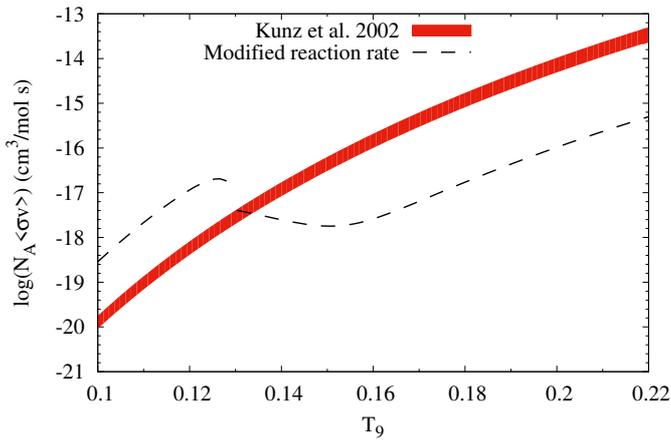}
  \caption{$\rm^{12}C+\alpha$ reaction rate at the CHeB temperatures, according to the work of \citet{2002ApJ...567..643K}
    (red thick line) together with the altered reaction rate necessary for mimicking the asteroseismic model for KIC 8626021 (dashed line).}
  \label{fig:rates-comp}
\end{figure}

In view of these findings, we computed the evolution of a progenitor
star from the ZAMS to the DB WD stage by altering significantly the
nuclear reaction rate for the purpose of mimicking the chemical
structure of KIC 08626021.
Because of the different temperatures at which CHeB and He-shell
burning proceed in the progenitor evolution, it is possible to alter
the $^{12}$C$+\alpha$ reaction rate to simultaneously reproduce the
large central O abundance and the existence of a C buffer derived by
\citet{2018Natur.554...73G}. Namely, we have been able to reproduce
the high central abundance for $^{16}$O ($\sim 82$\% by mass) by
enhancing the $\rm ^{12}C+\alpha$ reaction rate during the CHeB
phase---up to 10 times larger than the highest value predicted by
\citet{2002ApJ...567..643K}---for $T \lesssim 0.13 \times 10^9$
K. Beyond the core, we manage to form a C mantle in the top of the
CO-core by reducing the generation of O in the outward moving He
burning shell, during post-CHeB evolution. To do this, we find it
necessary to decrease the reaction rate in the range
$0.13 \times 10^9 \lesssim T$ by about 100 to 1000 times from that
predicted by \citet{2002ApJ...567..643K}. Only in this way, we find a
C dominated buffer ($\sim 90\%$), with a small amount of O.
In Fig. \ref{fig:rates-comp} we compare the standard
$\rm^{12}C+\alpha$ reaction rate with its current uncertainties 
  --$\pm 30\%$ of relative uncertainty -- at the CHeB temperatures
(red thick line) together with the altered reaction rate necessary for
reproducing the asteroseismic model for KIC 8626021 (dashed line). It
is clear that the uncertainty in the ${\rm ^{12}C} + \alpha$ reaction
rate cannot be invoked to produce the high O abundance in the core
and the almost C-pure buffer of the \citet{2018Natur.554...73G} WD
profile.

We also explore possible uncertainties in the Coulomb screening
factors could lead to the formation of such features in the chemical
structure.  The screening corrections are applied as a multiplicative
factor of the form $\exp{f}$ to the nuclear reaction rates, where the
factor $f$ depends upon the charge of the nucleus taking part in the
reactions. Up to now, all the recipes of screening corrections are
derived within certain assumptions
\citep{1973ApJ...181..439D,1973ApJ...181..457G,1982ApJ...258..696W}. However,
we need to keep in mind that any change in the screening correction of
a particular reaction is not going to affect only that reaction, but
possibly also every nuclear reaction where one of the same nucleus are
involved. This inhibits us from doing extreme changes in the screening
factors. In particular, as discussed in the previous paragraph, the
${\rm ^{12}C}+\alpha$ reaction rate is the major responsible for
setting the interior profile of a WD, both during the CHeB and the
He-shell burning phases.
If we interpret that this change is due
to the uncertainty in the screening factor, we need $\exp{f}$ to be
more than one order of magnitude higher than the one calculated by our
code (taken from \citealt{1973ApJ...181..457G} and
\citealt{1982ApJ...258..696W}), for temperatures up to $T_9 \sim 0.13$
and lower for temperatures $T_9\gtrsim 0.13$ (two order of magnitude
lower for $T_9\gtrsim 0.16$). These extreme changes in the screening
of the ${\rm ^{12}C}+\alpha$ reaction rate should affect other
screening factors for reactions involving C, He, or both (or even
other isotopes, due to the temperature dependence of the change), most
probably changing drastically other parts of stellar evolution that
are well constrained.

\subsection{Thermal pulses on the AGB}
\label{sect:tp}

\begin{figure}
  \includegraphics[width=0.35\textwidth, angle=270]{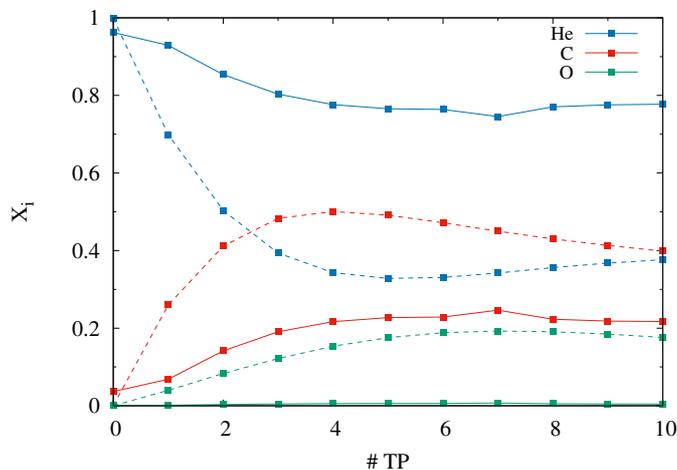}
  \caption{Intershell abundances of $\rm ^{4}He,^{12}C$ and
    $^{16}$O during the evolution on the thermally pulsing AGB
    phase. Dashed (solid) lines refers to the model in which (no) OV is
    considered in this stage.}
  \label{fig:he-c-o-pt}
\end{figure}  

Three main features in the chemical structure of the asteroseismic
model can be connected with physical processes occurring at the TP-AGB
phase: the CHe-plateau located beyond the C buffer, the total content
of He and the size of the degenerate core. The C-He plateau is the
result of the short-lived convective episodes occurring at the
He-burning shell, which dredge up C and shape the  flattened
profile.  The amount of O, C and He left at this intershell region
depends on the strength of the CBM at the border of the pulse-driven
convection zone, where $f\sim 0.0075$ reproduces reasonably well both
the initial to final mass relation and the abundances of PG1159 stars
\citep{2016A&A...588A..25M}.  Particularly, the intershell abundances
derived for the asteroseismic model ($\rm C \sim 80\%$, see lower
panel of Fig.\ref{fig:perfil-DB-0548}) disagree with both the results
from \citet{2000A&A...360..952H} and our computations, as seen from
Fig. \ref{fig:he-c-o-pt}. There we show the intershell abundances
resulting from the computations of a $M_{\rm ZAMS}=1.5M_{\odot}$
(final CO-core mass $M_{\rm CO}\sim 0.58 M_{\odot}$) model adopting
extreme values for the overshooting parameter $f=0$ and $f=0.0174$
(solid and dashed lines, respectively) in terms of the number of
thermal pulses experienced by the star on the AGB. These extreme
values of $f$ widely cover the current overshooting uncertainties
during the TP-AGB phase. We find that the maximum amount of $^{12}$C
in the intershell region ($\sim$ 50\%) occurs at the very first
thermal pulses of the model with $f=0.0174$, and this abundance is
still far below the $^{12}$C abundances derived for KIC 08626021.

The low total content of He of the asteroseismological model could be explained if the star experiences a long lived TP-AGB phase, i.e., if the star experiences a large number of thermal pulses. We find that it is possible to reduce the total He content of the star from $1.7\times 10^{-1}$ to $1\times 10^{-2}M_{\odot}$ in the  course of 10 thermal pulses. A total He content of
$10^{-4}M_{\odot}$, as found for KIC 08626021, would be possible if the
star experiences more than 30 thermal pulses. But in this case,  the growth of the core would largely exceed the mass derived for the
asteroseismic model. Therefore, it is not possible to find, in this
context, a model with an extremely low content of
He for an average mass WD.  Such He content is found for
ultra-massive WDs \citep{2019A&A...625A..87C}. Similar results are found by \citet[][see Figs. 9 and 10]{2006MNRAS.371..263L} where the authors find a final He content of $\sim 6\times 10^{-4}M_{\odot}$ for a WD  of $1.05M_{\odot}$.

A drawback arises when attempting to reproduce the intershell abundances  and the low content of He for the same model. The inclusion of CBM during the TP-AGB phase favors the occurrence of third dredge up episodes that prevent the core from growing and leads to the C-enrichment of the surface layers. The pollution of the stellar surface with C  drives strong winds, with the result of an earlier departure from the TP-AGB. This is in contrast with a long lived TP-AGB phase needed for the depletion of He to values close to $\sim 1\times 10^{-4}M_{\odot}$.

In light of the previous discussion, it appears difficult that the physical processes operative at the TP-AGB phase within their respective uncertainties,  could lead to the scenario in which an average-mass WD is formed with a C-rich intershell region simultaneously with a very low He content.

\section{Summary and conclusions}
\label{sect:summary}

\citet{2018Natur.554...73G} have performed for the first time an
extremely precise asteroseismological study of KIC 8626021, a DBV star
extensively monitored by the {\it Kepler} mission. The authors have
been able to find an asteroseismic model with an unprecedented
precision in their pulsation period match. This pave the way to dig
into the physical processes that lead to the formation of WD
stars.  The chemical structure derived from \citet{2018Natur.554...73G}
from their asteroseismological analysis for KIC 8626021 is not in
agreement with what is expected for a DB white dwarf star in terms of
the widely accepted formation channels, thus posing a challenge to the
theory of white-dwarf formation.  In this work, we have explored to
what extent both microphysics (diffusion processes and nuclear
reaction rates) and macrophysics (convective boundary mixing,
semiconvection) processes should be modified in order to reproduce the
chemical structure asteroseismologically derived for the DB pulsating
WD KIC 8626021 by \citet{2018Natur.554...73G}.  To this end, we
computed the evolution of progenitor stars from the ZAMS to the DBV
domain with final masses $M_{\rm WD}\sim$ 0.58 $M_{\odot}$.  As a
first step we explored the extent of the convective boundaries during
the CHeB phase in order to reproduce the mass of the large central
homogeneous part of the core. Based on the arguments presented by
\citet{2018Natur.554...73G}, we explored the impact of the extension
of the convective core on the final size of the homogeneous central
part of the WD core by enhancing the overshooting up to 5 times the
standard value. Even with such a large extension of the convective
boundary, our models are unable to develop a homogeneous central part
of the core of $M \sim 0.45 M_{\odot}$. We also explored the
efficiency of the diffusion processes acting during the WD cooling
path, in order to mimic the C buffer at the top of the core. We
evolved a $\sim$ 0.57$M_{\odot}$ WD model from $\sim 200000$ K to the
DB phase ($\sim 30000$ K) in which we varied the efficiency of the
diffusion processes from 0.01 to 100 times the standard value. We
found that diffusion is unable to create the C buffer at the top of
the core within a reasonable timescale.  Additionally, we found that
the thin He envelope that characterizes the asteroseismic model could
take place if the star cool down 50 to 100 times faster than normal DB
stars. 

In view of these findings and assuming  that  diffusion  is  the  only  process  able to modify the chemical structure during WD stage, these chemical features located so deep into the interior of the star, should be created during the evolution of the progenitor star, during the CHeB and AGB phases. We computed the complete evolution of a progenitor star in which
we altered the
$\rm^{12}C(\alpha,\gamma)^{16}O$ nuclear reaction rate during the whole
evolution. By modifying the nuclear reaction rate far beyond the extreme values predicted by \citet{2002ApJ...567..643K}, we have been able to reproduce a C buffer at top of an
O-dominated core. In particular, these features are only achieved if we enhance the nuclear reaction rate up to 10
times for $T<0.13\times 10^9$ K, and lowering down to 100--1000
times for $T>0.13\times 10^9$ K, which clearly lies outside the values suggested by laboratory determinations, including their uncertainties. In addition, we discarded that such features in the chemical structure could be reproduced by altering the screening factors within their uncertainties. 
We discussed the presence of the C-rich CHe-plateau and the low He content in the whole asteroseismic model. We found that a long-lived TP-AGB phase could be a possible scenario for the formation of a low-He content star, but we  envisage that such He content will be possible for WDs with $M_{\star} \approx 1.05 M_{\odot}$. This result is in contrast with the inclusion of CBM at the TP-AGB phase, a necessary ingredient to reproduce the intershell abundances. 

The results found in this work suggest that the asteroseismic model
for KIC 8626021 found by \citet{2018Natur.554...73G} is difficult to
reconcile with  our current understanding of the standard
evolutionary scenario for the formation of WDs. Further
  investigations are needed to understand the origin of this discrepancy.

In closing, it is appropriate to comment that
\citet{2018ApJ...867L..30T} have shown that even the very feeble
impact of neutrino emission on the mechanical structure of the WDs is
enough to alter low-order $g$-mode frequencies by about $70 \mu$Hz,
having a sizeable impact on WD mass, radius, and central O mass
fraction. Numerical experiments on our full evolutionary models show
that the presence of small chemical details left by previous evolution
(e.g. the small O bump at $-q \sim 2$, see the upper panel of
Fig. \ref{fig:perfil-DB-0548}) can alter low-order $g$-mode periods by
$\sim 0.1$ s ($\sim 10^7 \mu$Hz).

\begin{acknowledgements}
  Part of this work was supported by AGENCIA through the Programa de
  Modernizaci\'on Tecnol\'ogica BID 1728/OC-AR, and by the PIP
  112-200801-00940 grant from CONICET.  This research has made use of
  NASA's Astrophysics Data System.
\end{acknowledgements}

\bibliographystyle{aa}
\bibliography{paper-cbuffer}

\end{document}